\documentclass[12pt]{article}%
\pdfoutput=1
\usepackage[nosort]{cite}
\usepackage{graphicx}
\usepackage{multicol}
\usepackage{amsfonts}
\usepackage{amssymb}
\usepackage{amsmath}
\usepackage{singleauthor}
\usepackage{afterpage}
\usepackage{setspace}
\usepackage{verbatim}
\usepackage{slashed}
\usepackage{color}
\usepackage{longtable}
\usepackage{float}
\usepackage{subcaption}
\usepackage{epsfig}
\usepackage{bm}
\usepackage{enumerate}
\usepackage{epstopdf}
\usepackage[enableskew, vcentermath]{youngtab}
\usepackage{adjustbox}
\newsavebox{\mysavebox}

\usepackage[margin=1in]{geometry}
\usepackage{titletoc}%
\usepackage{hyperref}

\newcommand{\ba}{\begin{eqnarray}}
\newcommand{\ea}{\end{eqnarray}}

\newcommand{\Mp}{M_{\rm Pl}}

\newcommand{\Ntot}{N_{{\rm tot}}}

\newcommand{\be}{\begin{equation}}
\newcommand{\ee}{\end{equation}}

\makeatletter \@addtoreset{equation}{section} \makeatother

\begin{document}

\date{\today}

\title{Conditions for (No) Eternal Inflation}

\institution{IAS}{\centerline{School of Natural Sciences, Institute for Advanced Study, Princeton, NJ 08540, USA}}

\authors{Tom Rudelius\worksat{\IAS}\footnote{e-mail: {\tt rudelius@ias.edu}}}

\abstract{We construct analytic and numerical solutions of the Fokker-Planck equation that arises in the context of stochastic inflation. We use these solutions to derive necessary conditions for eternal inflation on the higher derivatives of the scalar field potential and examine the prospects for eternal inflation in a variety of popular models. We note similarities between the conditions needed to avoid eternal inflation and several recently-proposed Swampland criteria, which leads us to speculate on the possibility that the de Sitter Swampland conjectures should be viewed as approximate consequences of a No Eternal Inflation principle.
}

\maketitle

\tableofcontents

\enlargethispage{\baselineskip}

\setcounter{tocdepth}{2}

\newpage

\section{Introduction}\label{sec:INTRO}

Standard lore holds that once cosmic inflation begins, it will never end \cite{Vilenkin:1999pi, Linde:2007fr, Guth:2007ng, Baumann:2009ds}. Instead, inflation \cite{Guth:1980zm,Albrecht:1982wi,Linde:1981mu} produces an infinite collection of pocket universes, which develop a fractal structure \cite{Aryal:1987vn, Linde:1993xx, Winitzki:2001np}. These pocket universes ``populate'' the vacua in the Landscape of string theory which, combined with anthropic selection effects, alleviates the fine-tuning problem of the cosmological constant \cite{Weinberg:1988cp}.

However, there are a couple of important gaps in the standard tale of eternal inflation. First off, much of the literature on eternal inflation was developed in the context of one or two simple large-field models of inflation, ignoring quantum gravitational corrections that are expected to arise in such models. It is important to study the prospects for eternal inflation in more general models, both small-field and large-field, and to understand how quantum gravitational effects are likely to modify such prospects.

Secondly, eternal inflation has been historically studied in the context of phenomenologically viable models of primordial inflation. This is a perfectly reasonable thing to do, since CMB measurements can actually give us bona fide experimental data about the inflationary potential over a limited range in field space. In recent times, however, there has been a flurry of interest in characterizing and constraining scalar field potentials in string theory at \emph{generic} points in field space (see e.g. \cite{Obied:2018sgi, Danielsson:2018qpa, Denef:2018etk, Roupec:2018mbn, Danielsson:2018ztv, Murayama:2018lie, Choi:2018rze, Andriot:2018wzk, Conlon:2018eyr, Garg:2018reu, Ooguri:2018wrx, Hebecker:2018vxz, Andriot:2018mav, Gonzalo:2019gjp}), regardless of whether or not these potentials are experimentally relevant for particle physics and cosmology in our universe. This motivates us to consider conditions for eternal inflation at generic points in field space, even those that are experimentally excluded from describing primordial inflation in the early stages of own universe.

Our analysis proceeds in several steps. First, in \S\ref{sec:FP}, we derive analytic, Gaussian solutions to the Fokker-Planck equation that describes the stochastic evolution of scalar fields in a linear or quadratic potential with one or more scalar fields. Next, in \S\ref{sec:CONDITIONS}, we use these solutions to compare the exponential decay of the probability of inflation with the exponential growth of the universe during inflation. When the growth proceeds faster than decay, eternal inflation occurs. In this way, we derive necessary conditions for eternal inflation on the first and second derivatives of the potential, which in a single-field model take the form
\begin{equation}
\frac{|V'|}{V^{3/2}} < \frac{\sqrt{2}}{2\pi} \frac{1}{M_{\rm Pl}^3}\,,~~~~ - \frac{V''}{V} < \frac{3}{M_{\rm Pl}^2}.
\label{eq:introeq}
\end{equation}
Using a numerical analysis, we compute further necessary conditions for eternal inflation on the $p$th derivative of the potential for $p > 2$, which take the form
\begin{equation}
\left[ - \sgn(\partial^p V) \right]^{p+1} \frac{|\partial^p V|}{V^{(4-p)/2}} <  \mathcal{N}_p  \Mp^{p-4}\,,~~~p>2,
\label{eq:introeq2}
\end{equation}
with $\mathcal{N}_p \gg 1$. For a metastable vacuum, all of these derivative conditions are satisfied, but eternal inflation further requires the vacuum decay rate per unit volume $\Gamma$ to be smaller than the Hubble expansion rate,
\begin{equation}
\frac{\Gamma}{H^4} < \frac{9}{4 \pi}.
\label{eq:introeq3}
\end{equation}

In \S\ref{sec:PRIMORDIAL}, we apply these conditions to a number of popular models of primordial inflation to see if/when they become eternal. We find that some models (e.g. power-law inflation, inflection point inflation) are frequently not eternal, whereas other models (e.g. hilltop inflation, natural inflation) are all but guaranteed to be eternal in order to agree with observation.

Our analysis in \S\ref{sec:CONDITIONS} and \S\ref{sec:PRIMORDIAL} extends the work of \cite{Creminelli:2008es}, which derived precisely the inequality in the left-hand side of (\ref{eq:introeq}). By considering hilltop potentials, we further derive the constraints in the right-hand side of (\ref{eq:introeq}) and (\ref{eq:introeq2}). Our hilltop analysis complements the work of \cite{Barenboim:2016mmw, Kinney:2018kew, Brahma:2019iyy, Dvali:2018jhn}, and there is a good deal of overlap between our results and theirs. However, by solving the Fokker-Planck equation analytically and numerically, we take a somewhat more rigorous approach in contrast to the more physically-intuitive approaches considered in those works. As a result, our bounds in (\ref{eq:introeq}) differ from theirs by $O(1)$ factors, and our bounds in (\ref{eq:introeq2}) differ from theirs by rather large factors, though our qualitative conclusions are unchanged. Our approach also admits a straightforward generalization to theories with multiple scalar fields.

In \S\ref{sec:SWAMPLAND}, our discussion becomes more exploratory. We point out similarities between the conditions needed to avoid eternal inflation and the constraints on scalar field potentials proposed in several ``de Sitter Swampland conjectures'' \cite{Obied:2018sgi, Garg:2018reu, Ooguri:2018wrx, Andriot:2018mav}. These similarities could be mere coincidences, or there may be some other deep, fundamental principle of quantum gravity that explains why the de Sitter conjectures are true, by which eternal inflation is accidentally outlawed as a corollary (as previously noted in \cite{Kinney:2018kew, Brahma:2019iyy, Matsui:2018bsy, Dimopoulos:2018upl}). But a more intriguing possibility is that the condition of No Eternal Inflation is \emph{itself} the deep, fundamental principle that explains why the de Sitter Swampland criteria should (approximately) hold true: if eternal inflation is incompatible with quantum gravity, for some presently unknown reason, it would imply that scalar field potentials in quantum gravity must violate at least one of the conditions (\ref{eq:introeq})-(\ref{eq:introeq3}), which would manifest in bounds that are very similar to those proposed in various de Sitter Swampland conjectures. This possibility is appealing in that it offers some sort of physical motivation for these conjectures, which is lacking at present. At the same time, it is easier to avoid eternal inflation than to satisfy the bounds imposed by these conjectures (e.g. metastable de Sitter vacua are allowed by the former but not the latter), which means that the restrictive constraints on phenomenology and string theory model-building from these conjectures could be relaxed by exchanging them for a No Eternal Inflation principle. Still, the constraints on scalar field potentials needed to avoid eternal inflation are strong enough that such a principle could be falsified by a reliable string theory construction of de Sitter critical points with sufficiently-light tachyons. We ponder the possible consequences and motivations of a No Eternal Inflation principle in quantum gravity, noting that such a principle would come with some attractive consequences as well as some ugly ones. Finally, we conclude our discussion in \S\ref{sec:CONC} with a brief summary of our results and a list of some open questions.

\section{Analytic Solutions to the Fokker-Planck Equation}\label{sec:FP}

The study of inflation begins by considering a scalar field theory in a quasi-de Sitter background,
\begin{equation}
S = \int d^4x \sqrt{-g} \left[\frac{1}{2} R +\frac{1}{2} g^{\mu\nu} \partial_\mu \phi \partial_\nu \phi - V(\phi) \right],
\end{equation}
with
\begin{equation}
ds^2 = -dt^2 + e^{2 H t}d\vec{x}^2.
\end{equation}
Restricting to the case of a homogenous scalar field $\phi(t, \vec{x}) := \phi(t)$, the dynamical equations describing the evolution of the scalar field and the background geometry are given by
\begin{equation}
\ddot \phi + 3 H \dot \phi + \frac{\partial V}{\partial \phi} = 0\,,~~~H^2  \Mp^2 = \frac{1}{3} \left(\frac{1}{2} \dot \phi^2 + V(\phi) \right).
\end{equation}
In the ``slow-roll'' approximation, these become
\begin{equation}
 3 H \dot \phi + \frac{\partial V}{\partial \phi} \approx 0\,,~~~H^2  \Mp^2 = \frac{1}{3} V(\phi).
\end{equation}
The standard treatment of eternal inflation proceeds by writing the scalar field as the sum of a long-wavelength classical background and a short-wavelength quantum field,
\begin{equation}
\phi(t,\vec{x}) = \phi_{cl}(t,\vec{x}) + \delta \phi (t,\vec{x}).
\end{equation}
The action for the quantum fluctuations is quadratic, so the fluctuations will be Gaussian. These fluctuations are averaged over a Hubble volume by defining a smeared field \cite{Bousso:2006ge},
\begin{equation}
\delta \phi_H(t) = \int \frac{d^3\vec{k}}{(2\pi)^3} \sigma_t(\vec{k}) \delta \phi_{\vec{k}} (t),
\end{equation}
with $\delta \phi_{\vec{k}}$ a Fourier mode of $\delta \phi$ and $\sigma_t(\vec{k})$ a smearing function that corresponds to averaging over one Hubble volume at a time $t$. The average size of these Gaussian fluctuations is given by \cite{Vilenkin:1982wt, Linde:1982uu, Starobinsky:1982ee}
\begin{equation}
\langle [\delta \phi_{H}(t) - \delta \phi_{H}(0)]^2  \rangle = \left( \frac{H}{2 \pi} \right)^2 H t,
\end{equation}
which is often stated in terms of the typical quantum fluctuation $\delta \phi_q$ over a Hubble time $t = H^{-1}$ as
\begin{equation}
\delta \phi_q = \frac{H}{2 \pi}.
\end{equation}
Once these quantum fluctuations exit the horizon, they decohere and behave classically \cite{Kiefer:1998qe}. Thus, assuming a large number of $e$-foldings, these quantum fluctuations can modeled as classical, Gaussian noise term added to the equation of motion for $\phi$,
\begin{equation}
3 H \dot \phi +  \frac{\partial V}{\partial \phi} = N(t),
\end{equation}
with $N(t)$ a Gaussian noise term, which induces a random walk of the field $\phi$ in the potential $V(\phi)$. Thus, in an infinitesimal time $\delta t$, $\phi$ will vary according to
\begin{equation}
\delta \phi = - \frac{1}{3 H} V'(\phi) \delta t + \delta \phi_q(\delta t),~~~\delta \phi_q(\delta t) \sim \mathcal{N}(0, H^3 \delta t/(2 \pi)^2).
\end{equation}
Here, the first term on the right-hand side represents the classical evolution of $\phi$ in the potential $V(\phi)$, and the second term represents the stochastic Gaussian noise due to quantum fluctuations.\footnote{It is worth noting that recent work in quantum cosmology has called into question this standard semiclassical treatment of eternal inflation, arguing instead that eternal inflation must be addressed within a genuinely quantum gravitational regime \cite{Hawking:2017wrd, Vreys:2017fpr}. In our analysis here and below, we simply assume the standard scenario.}

The evolution of the probability distribution of $\phi$ is then described by a Fokker-Planck equation \cite{Starobinsky:1986fx, Rey:1986zk, Linde:1991sk}. Generalizing now to a theory of multiple scalar fields $\bm{\phi} = \phi^i, i = 1,...,N$, this equation takes the form
\begin{equation}
\dot P[\bm{\phi},t] = \frac{1}{2} \left( \frac{H^3}{4 \pi^2} \right) \partial_i \partial^i P[\bm{\phi},t] + \frac{1}{3 H}  \partial_i \Big( (\partial^i V(\bm{\phi})) P[\bm{\phi},t] \Big),
\label{eq:FP}
\end{equation}
where $\partial_i := \partial/\partial \phi^i$, $\dot P := \partial P/\partial t$, and we have assumed a trivial metric on field space, $g_{ij} = \delta_{ij}$.

This equation is difficult to solve for a general potential $V(\bm{\phi})$, but it can be solved for a sum-separable potential:
\begin{equation}
V(\bm{\phi}) = V_0 + \sum_{i=1}^N V_i(\phi^i),
\end{equation}
with $V_i(\phi_i)$ linear or quadratic, under the assumption that $H$ is constant, i.e.
\begin{equation}
H^2 M_{\rm{Pl}}^2 = \frac{V_0}{3}.
\label{eq:Freedman}
\end{equation}
Under these assumptions, the solution takes a simple multivariate Gaussian form:
\begin{equation}
P[\bm{\phi},t] = \prod_{i=1}^N P_i [\phi_i,t],
\end{equation}
with 
\begin{equation}
 P_i [\phi_i,t] = \frac{1}{ \sigma_i(t)\sqrt{2 \pi}} \exp\left[-  \frac{(\phi_i - \mu_i(t))^2}{2 \sigma_i(t)^2}\right]
 \label{eq:Gaussian}
\end{equation}
It suffices to consider each field separately, so for the remainder of this section, we specialize to the case of a single scalar field $\phi$ with a potential $V(\phi)$. Our Gaussian ansatz suffices for four different classes of potential: constant, linear, quadratic, and tachyonic. We consider each of them in turn.\footnote{A closely related analysis has been previously carried out in \cite{Rey:1986zk}.}

\vspace{.2cm}

\noindent
\underline{\bf Case 1: Free Massless Field ($V(\phi) = V_0$)}

\vspace{.1cm}
\noindent
For a constant potential $V(\phi) = V_0$, (\ref{eq:FP}) is solved by a Gaussian distribution (\ref{eq:Gaussian}) with
\begin{equation}
\mu(t) = 0\,,~~~~\sigma^2(t) = \frac{H^3}{4 \pi^2} t.
\label{eq:freemassless}
\end{equation}
We see that a delta-function distribution initially centered at $\phi = 0$ will remain centered at $\phi=0$ for all time, but it will spread out by an amount $\sigma(t = H^{-1}) = H/2\pi$ after a Hubble time. This represents the standard ``Hubble-sized'' quantum fluctuations that are well-known in the context of inflation, famously imprinting in the CMB and ultimately seeding the observed large-scale structure. One expects that this analysis will carry over to good approximation for any sufficiently-flat potential, like the type required for slow-roll inflation.

\vspace{.2cm}

\noindent
\underline{\bf Case 2: Linear Potential ($V(\phi) = V_0 - \alpha \phi$)}

\vspace{.1cm}
\noindent
For a linear potential $V(\phi) = V_0 - \alpha \phi$, (\ref{eq:FP}) is again solved by a Gaussian distribution (\ref{eq:Gaussian}) with
\begin{equation}
\mu(t) = \frac{\alpha}{3 H} t  \,,~~~~\sigma^2(t) = \frac{H^3}{4 \pi^2} t.
\label{eq:linearevolution}
\end{equation}
Each of these terms admits a very simple explanation: the time-dependence of $\mu(t)$ is explained by the classical rolling of the field in the linear potential, which is governed by the slow-roll equation of motion
\begin{equation}
3 H \dot \phi = - \frac{\partial V}{\partial \phi} = \alpha.
\end{equation}
The time-dependence of $\sigma^2(t)$, on the other hand, is due purely to Hubble-sized quantum fluctuations, and indeed it precisely matches the result in the free massless case.

\vspace{.2cm}

\noindent
\underline{\bf Case 3: Free Massive Field ($V(\phi) = V_0 + \frac{1}{2} m^2 \phi^2$)}

\vspace{.1cm}
\noindent
For a linear potential $V(\phi) = V_0 + \frac{1}{2} m^2 \phi^2$, (\ref{eq:FP}) is solved by a Gaussian distribution (\ref{eq:Gaussian}) with
\begin{equation}
\mu(t) = 0 \,,~~~~\sigma^2(t) = \frac{3 H^4}{8 \pi^2 m^2}\left(1 - \exp\left[  -\frac{2 m^2}{3 H}t \right] \right).
\end{equation}
As expected by symmetry, the central value of the Gaussian distribution remains fixed at $\phi =0$ for all time. In the limit $m/H \rightarrow 0$, we may expand the exponential to linear order to recover the formula for $\sigma^2(t)$ for the free massless field case in (\ref{eq:freemassless}). However, when $t \gtrsim H/m^2$, we see that the spreading of the distribution stops: the mass term prevents fluctuations larger than $O(H^2/m)$.

\vspace{.2cm}

\noindent
\underline{\bf Case 4: Tachyonic Field ($V(\phi) = V_0 - \frac{1}{2} m^2 \phi^2$)}

\vspace{.1cm}
\noindent
Finally, we consider a tachyonic potential $V(\phi) = V_0 - \frac{1}{2} m^2 \phi^2$. Now, (\ref{eq:FP}) is solved by a Gaussian distribution (\ref{eq:Gaussian}) with
\begin{equation}
\mu(t) = 0 \,,~~~~\sigma^2(t) = \frac{3 H^4}{8 \pi^2 m^2}\left(-1 + \exp\left[  \frac{2 m^2}{3 H}t \right] \right).
\label{eq:tachyonevolution}
\end{equation}
Once again, the distribution stays centered at $\phi=0$, and for $m/H \ll 1$, we may expand the exponential to linear order to recover the formula for $\sigma^2(t)$ for the free massless field case in (\ref{eq:freemassless}). Now, however, the tachyonic instability leads to an exponential \emph{growth} of the standard deviation $\sigma$ on time scales larger than $H/m^2$. This occurs, for instance, in spontaneous symmetry breaking, as the field quickly decays from the hilltop at $\phi = 0$ and settles into a vacuum with nonzero expectation value.


\section{Conditions for Eternal Inflation}\label{sec:CONDITIONS}

Now that we have understood the solutions for the Fokker-Planck equation for a variety of simple potentials, we are in a position to answer our original question: under what circumstances will eternal inflation occur?

To answer this, we must first develop an intuitive understanding for how inflation becomes eternal. For purposes of illustration, it is useful to consider a linear potential, $V(\phi) = V_0  - \alpha \phi$. In this potential, inflation occurs provided 
\begin{equation}
\epsilon_V(\phi) := \frac{M_{\rm Pl}^2}{2} \left( \frac{\alpha}{V(\phi)} \right)^2 \ll 1.
\label{eq:slowrollcond}
\end{equation}
For $\phi \leq 0$, this holds provided $\alpha M_p \ll V_0$. But as $\phi$ rolls down its slope, $V(\phi)$ will shrink and $\epsilon_V$ will grow, violating (\ref{eq:slowrollcond}) by the time $\phi = V_0/\alpha - M_{\rm Pl}/\sqrt{2}$.

So, suppose we want to know the probability that $\phi > \phi_{c}:=V_0/\alpha - M_{\rm Pl}/\sqrt{2}$ after a time $t$, assuming that $\phi$ starts from rest at $\phi =0$ at time $t=0$. From (\ref{eq:linearevolution}), we see that the Gaussian describing the location of $\phi$ shifts linearly with time, whereas its standard deviation grows only as $\sigma \sim \sqrt{t}$. The upshot of this is that the probability,
\begin{equation}
\text{Pr}[\phi> \phi_c, t] = \int_{-\infty}^{\phi_c} d\phi \,P[\phi,t]
\end{equation}
will tend to 0 in the $t \rightarrow \infty$ limit. It seems that inflation does not last forever: the field will leave the inflationary regime $\phi > \phi_c$ with probability 1. How then could inflation be eternal?

The key is that the evolution of $\phi$ here actually competes with another effect: the expansion of the universe. Assuming an initial volume for the universe $U(t=0) =U_0$ and constant Hubble rate $H$, the volume of the universe after a time $t$ will be given by
\begin{equation}
U(t) = U_0 (a(t)/a_0)^3 = U_0 e^{3 H t}.
\end{equation}
Interpreting the probability Pr$[\phi> \phi_c, t]$ as the fraction of the total volume with $\phi > \phi_c$,\footnote{This interpretation seems to make sense in the context of eternal inflation, as the expansion will produce infinitely many Hubble-sized pocket universes out of causal contact, and we expect these universes will follow the Fokker-Planck distribution.} we find that the volume of space still inflating at time $t$ is given by
\begin{equation}
U(\phi> \phi_c, t) = \text{Pr}[\phi> \phi_c, t] \times U(t) =  \text{Pr}[\phi> \phi_c, t] \times U_0 e^{3 Ht}.
\end{equation}
Thus, even though Pr$[\phi> \phi_c, t]$ will decrease with $t$, inflation will still be eternal provided it shrinks at a rate slower than the Hubble expansion rate $\exp 3 Ht$. To derive necessary conditions for eternal inflation, therefore, we simply compute Pr$[\phi> \phi_c, t]$ and insist that it shrinks more slowly than the Hubble expansion rate.

Using the results of the previous section, we carry out this computation analytically for three relevant cases: 1) a linear potential, 2) a quadratic hilltop potential, and 3) a combination of the two. We then adopt a numerical approach to carry out the computation for more general hilltop potentials.

\subsection{Linear and Quadratic Hilltop Models}\label{ssec:LinearHilltop}

\vspace{.2cm}

\noindent
\underline{\bf Case 1: Linear Potential ($V(\phi) = V_0 - \alpha \phi$)}

\vspace{.1cm}
\noindent
In general, slow-roll inflation occurs when the first derivative of the potential $V'(\phi)$ is small relative to the potential itself, so $V$ can be approximated locally as a constant in $\phi$. In our linear model, this will happen for $\phi$ less than some critical value $\phi_c$, the precise value of which is not important for our purposes (as we will soon see). The probability $\text{Pr}[\phi> \phi_c, t]$ is given by 
\begin{equation}
\text{Pr}[\phi> \phi_c, t] = \int_{-\infty}^{\phi_c} d\phi \,P[\phi,t],
\end{equation}
where $P[\phi,t]$ is the probability density function for a Gaussian distribution, given in (\ref{eq:Gaussian}), with mean $\mu(t)$ and variance $\sigma^2(t)$ given by (\ref{eq:linearevolution}). The result is
\begin{equation}
\text{Pr}[\phi> \phi_c, t]  = \frac{1}{2} \text{erfc}\left[\frac{\mu(t)-\phi_c}{\sigma(t) \sqrt{2} } \right] =  \frac{1}{2} \text{erfc}\left[\frac{\frac{\alpha }{3 H}t-\phi_c}{ \frac{H}{2\pi}\sqrt{2Ht} } \right],
\end{equation}
with erfc the error function. For large $t$, this error function can be approximated as an exponential,
\begin{equation}
\text{Pr}[\phi> \phi_c, t] \approx C(t) \exp \left[ - \left( \frac{\frac{\alpha }{3 H}t-\phi_c}{ \frac{H}{2\pi}\sqrt{2Ht} } \right)^2 \right] \approx C(t) \exp \left[ -   \frac{4 \pi^2 \alpha^2}{18 H^5} t \right],
\label{eq:lasteq}
\end{equation}
with $C(t)$ a power-law in $t$. We see that indeed, the value of $\phi_c$ has dropped out, and the question of whether or not inflation is eternal hinges on whether or not this exponential decay beats the exponential expansion of the universe. In particular, eternal inflation happens when
\begin{equation}
3 H  > \frac{4 \pi^2 \alpha^2}{18 H^5}.
\end{equation} 
Exchanging $H$ for $V$ using $H^2 M_{\rm Pl}^2 = V/3$ and setting $\alpha = V'(\phi)$, this becomes
\begin{equation}
\frac{|V'|}{V^{3/2}} < \frac{\sqrt{2}}{2\pi} \frac{1}{M_{\rm Pl}^3}
\label{eq:lineareternal}
\end{equation}
This is a familiar expression in the context of eternal inflation. It can be thought of as the condition that the quantum fluctuations in a Hubble time, $\delta \phi_{q} = \sigma(t = H^{-1}) = H/2\pi$, dominate over the classical rolling in a Hubble time, $\delta \phi_{cl} = V'(\phi)/3H^2$. Of course, this condition will be satisfied for a linear potential when $\phi \rightarrow - \infty$, since $V(\phi)$ grows without bound in this limit. More realistically, however, we might expect that the linear approximation is valid over a smaller regime, in which case (\ref{eq:lineareternal}) may or may not be satisfied.

In a slightly more rigorous analysis, we might have allowed $H$ to depend on $\phi$, rather than treating it as a constant. We might also have added a barrier to account for the fact that inflation ends when $|V"|/V \sim \Mp^{-1}$, as was done in \cite{Creminelli:2008es}. However, these modifications would not affect our conclusions, which depend only on the argument of the exponential (\ref{eq:lasteq}). Indeed, the analysis of \cite{Creminelli:2008es} (with a barrier included) produced precisely the same bound (\ref{eq:lineareternal}) as our analysis, including the numerical coefficient on the right-hand side.

We do not need to consider multi-dimensional linear potentials, as we can always change our field basis to turn a sum of $N$ linear potentials into a single linear potential and $N-1$ massless free fields, the latter of which do not affect the probability of eternal inflation.

\vspace{.2cm}

\noindent
\underline{\bf Case 2: Quadratic Hilltop Potential ($V(\phi) = V_0 - \frac{1}{2} m^2 \phi^2$)}

\vspace{.1cm}
\noindent
For a quadratic hilltop potential, inflation occurs when $|\phi|$ is smaller than some critical value $\phi_c$. In light of our result for the linear case, we might (following \cite{Barenboim:2016mmw, Kinney:2018kew}) justifiably define $\phi_c$ to be the the value of $\phi$ at which $|V'| / V^{3/2} = 1/(2 \pi M_{\rm Pl}^3)$,
\begin{equation}
\phi_c := \frac{V_0^{3/2}}{2 \pi m \Mp^3},
\end{equation}
since for $|\phi| > \phi_c$ we know that the classical rolling of $\phi$ will dominate over the quantum fluctuations that keep it on top of the hill. However, for the purposes of our analysis, we do not need to be too dogmatic about the value of $\phi_c$: as in the linear case, its value will produce subleading effects that do not affect our conclusions.

We are interested in the probability that the field lies between $-\phi_c$ and $\phi_c$ after a time $t$. This is given by 
\begin{equation}
\text{Pr}[|\phi| < \phi_c, t] = \int_{-\phi_c}^{\phi_c} d\phi \,P[\phi,t],
\end{equation}
The probability density function is again a Gaussian, with mean and variance given by (\ref{eq:tachyonevolution}). As a result, we have
\begin{equation}
\text{Pr}[|\phi| < \phi_c, t]   = \text{erf}\left[\frac{\phi_c-\mu(t)}{\sigma(t) \sqrt{2}} \right] =  \text{erf}\left[ \frac{ 2 \pi m \phi_c }{  \sqrt{3} H^2 (-1+ \exp \frac{2 m^2 }{3 H}t)^{1/2} } \right], 
\end{equation}
As $t\rightarrow \infty$, the denominator blows up. The error function admits a Taylor expansion in this limit,
\begin{equation}
\text{erf}(x) = \frac{2}{\sqrt{\pi}} \frac{1}{x} + O\left(\frac{1}{x}\right)^3,
\end{equation}
so we may write
\begin{equation}
\text{Pr}[|\phi| < \phi_c, t]  \approx  \frac{2}{\sqrt{\pi}} \frac{2 \pi m \phi_c}{\sqrt{3} H^2} \exp\left[ - \frac{m^2}{3 H} t
\right]
\end{equation}
We see that indeed, the precise value of $\phi_c$ is irrelevant, and the important question for eternal inflation is whether or not this exponentially decaying probability beats the exponential expansion of the universe. Namely, eternal inflation occurs if
\begin{equation}
3 H > \frac{m^2}{3 H},
\end{equation}
Using $H^2 M_{\rm Pl}^2 = V/3$ and setting $-m^2 = V''$, we see that eternal inflation occurs if
\begin{equation}
\frac{V''}{V} > -\frac{3}{M_{\rm Pl}^2}.
\label{eq:hilltopeternal}
\end{equation}
This equation is not as familiar as the condition (\ref{eq:lineareternal}) derived in the linear case, but it has also been derived previously (up to $O(1)$ factors) in \cite{Barenboim:2016mmw, Kinney:2018kew, Dvali:2018jhn} by a slightly different logic. Namely, if one views the characteristic time for $\sigma(t)$ to grow to size $\phi_c$ as a lifetime for the field to exit the hilltop, then (\ref{eq:hilltopeternal}) follows up to $O(1)$ factors from demanding that this lifetime should be longer than the Hubble time $H^{-1}$, so that Hubble expansion occurs more quickly than $\phi$ exits the hilltop. It is comforting that these two distinct approaches lead to roughly the same bound, though not too surprising given that the only mass scales that show up in the Fokker-Planck equation are $m$ and $H$. The more important result here is that our linear and quadratic analyses produced an exponentially decaying probability that is \emph{linear} in time, so it competes directly with the exponential Hubble expansion.

This analysis may be generalized straightforwardly to the case of a multi-dimensional hilltop potential. We saw in \S\ref{sec:FP} that the probability density function separates into a multivariate normal distribution. Eternal inflation occurs when $|\bm{\phi}|$ lies inside some small ellipsoid $\mathcal{E}$ centered at the origin, but as in the single-field case, the precise size or shape of this region is unimportant: all that matters is exponential dependence of the standard deviation $\sigma_i(t)$. Namely, we have
\begin{equation}
\text{Pr}[\bm{\phi}(t) \in \mathcal{E}]  \sim \sigma_1(t) \cdot \sigma_2(t) \cdot ... \cdot \sigma_N(t) \sim \exp \left[ -\frac{t}{3H} \sum_{i=1}^N m_i^2 \right].
\end{equation}
Eternal inflation occurs when
\begin{equation}
3 H > \frac{1}{3 H} \sum_{i=1}^N m_i^2,
\end{equation}
or equivalently,
\begin{equation}
\frac{\partial_i \partial^i V(\bm{\phi})}{V(\bm{\phi})} > -\frac{3}{M_{\rm Pl}^2}.
\label{eq:multihilltopeternal}
\end{equation}
This equation holds in the case of a local maximum. In the case of a saddle point, the sum over $i$ should be taken over only the tachyonic directions, since the probability density function does not spread out significantly in the directions with positive mass term.

\vspace{.2cm}

\noindent
\underline{\bf Case 3: Combination of Linear Potential and Quadratic Hilltop Potential}

\vspace{.1cm}
\noindent
As our final case, we put cases 1 and 2 together and consider what happens when we have a single linear direction and $N-1$ tachyonic ones. In other words, 
\begin{align}
V(\bm{\phi}) = V_0 - \alpha_1 \phi^1 - \sum_{i=2}^N \frac{1}{2} m_i^2 \phi^i
\end{align}
Inflation will happen for $\phi_1 > \phi_{1,c}$, and $\phi^i$, $i=2,...,N$ contained in a small region $\mathcal{E}$ centered at the origin in these direction. Once again, the condition for eternal inflation is determined by the spreading of the probability density function with time, which is given by
\begin{align}
\text{Pr}[\phi^1(t) < \phi_c, \phi^{2,...,N}(t) \in \mathcal{E}]  &\sim \exp\left[ - \frac{\mu_1(t)^2}{2 \sigma_1(t)^2} \right] \sigma_2(t) \cdot \sigma_3(t) \cdot ... \cdot \sigma_N(t) \nonumber \\
&\sim \exp \left[  - \left( \frac{4 \pi^2 \alpha_1^2}{18 H^5} + \frac{1}{3H} \sum_{i=2}^N m_i^2\right) t \right].
\end{align}
This means that inflation will be eternal if
\begin{equation}
3 H > \frac{4 \pi^2 \alpha_1^2}{18 H^5} + \frac{1}{3H} \sum_{i=2}^N m_i^2,
\end{equation}
or
\begin{equation}
\frac{V}{\Mp^4} > \frac{4 \pi^2 \Mp^2 (\partial_1 V)^2}{2 V^2} + \frac{1}{3 \Mp^2} \sum_{i=2}^N \partial_i \partial_i V.
\label{eq:combined}
\end{equation}

\subsection{General Hilltop Models}\label{ssec:generalhill}

So far, we have considered conditions for eternal inflation in linear and quadratic hilltop models. What about more general hilltops of the form $V(\phi) = V_0  - \frac{1}{p} \alpha \phi^p$, for $p > 2$? These models cannot be analyzed analytically, because the probability distribution $P[\phi,t]$ cannot be described by a Gaussian for all time. Instead, we must turn to a numerical analysis.

To do this, we discretize the slow-roll equation of motion for $\phi$:
\begin{equation}
\phi_{n} = \phi_{n-1} - \frac{1}{3 H} V'(\phi_{n-1}) \delta t + \delta \phi_q(\delta t)\,,~~~\phi_0 = 0,
\end{equation}
with $\delta t$ the step size and the quantum fluctuation term $\delta \phi_q(\delta t)$ drawn from a Gaussian distribution with mean $0$ and variance $H^2/(2 \pi)^2 \delta t$. We then solve these equations numerically and study the probability that $\phi$ is still on the hilltop after $N$ time steps. In our analysis, we set $\delta t = 1/100 H^{-1}$ to avoid overshooting. Note that we also treat $H$ as a constant, which is a valid approximation at the top of the hill where our analysis is relevant.

\begin{figure}
\begin{center}
\includegraphics[width=80mm]{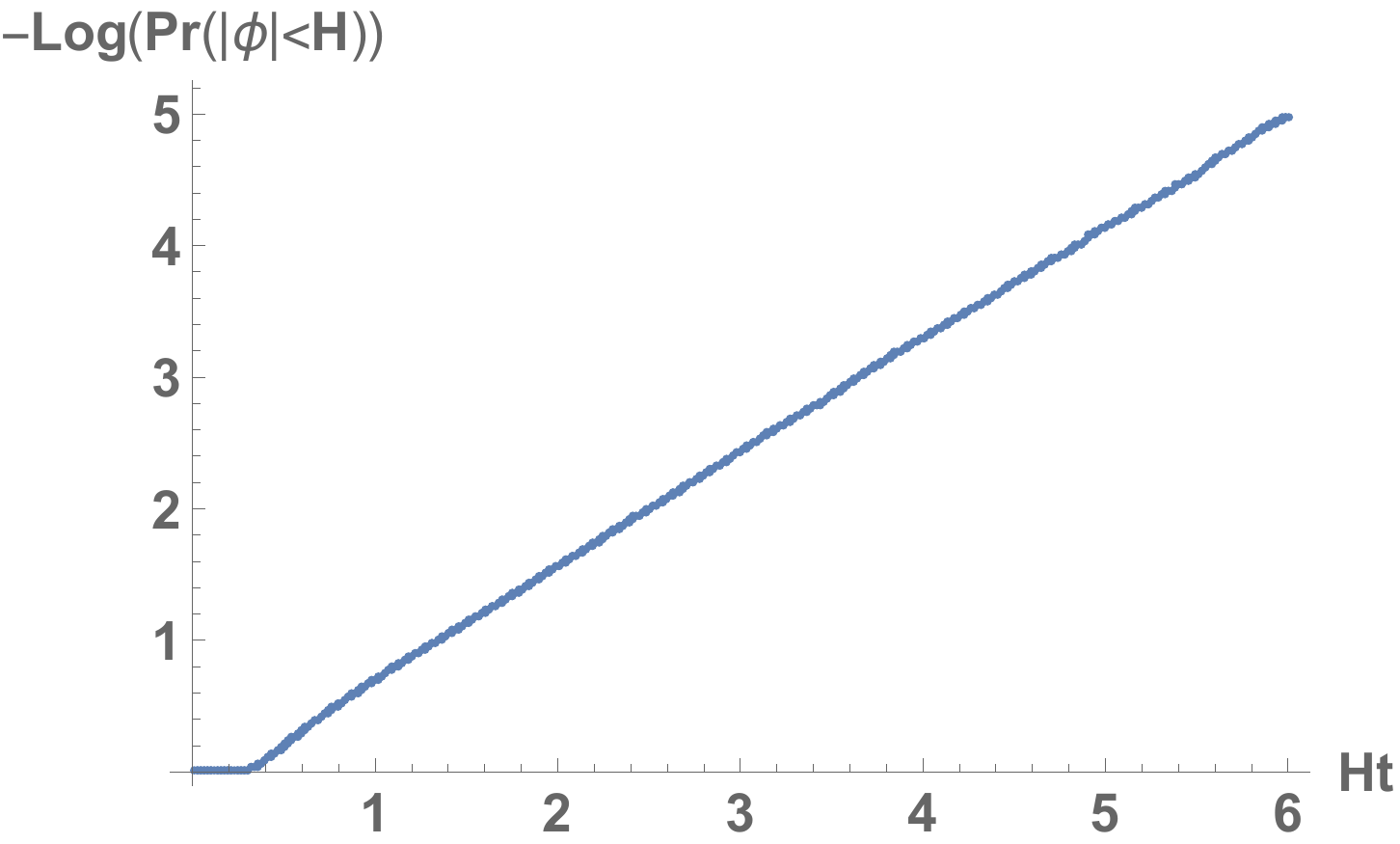}
\includegraphics[width=80mm]{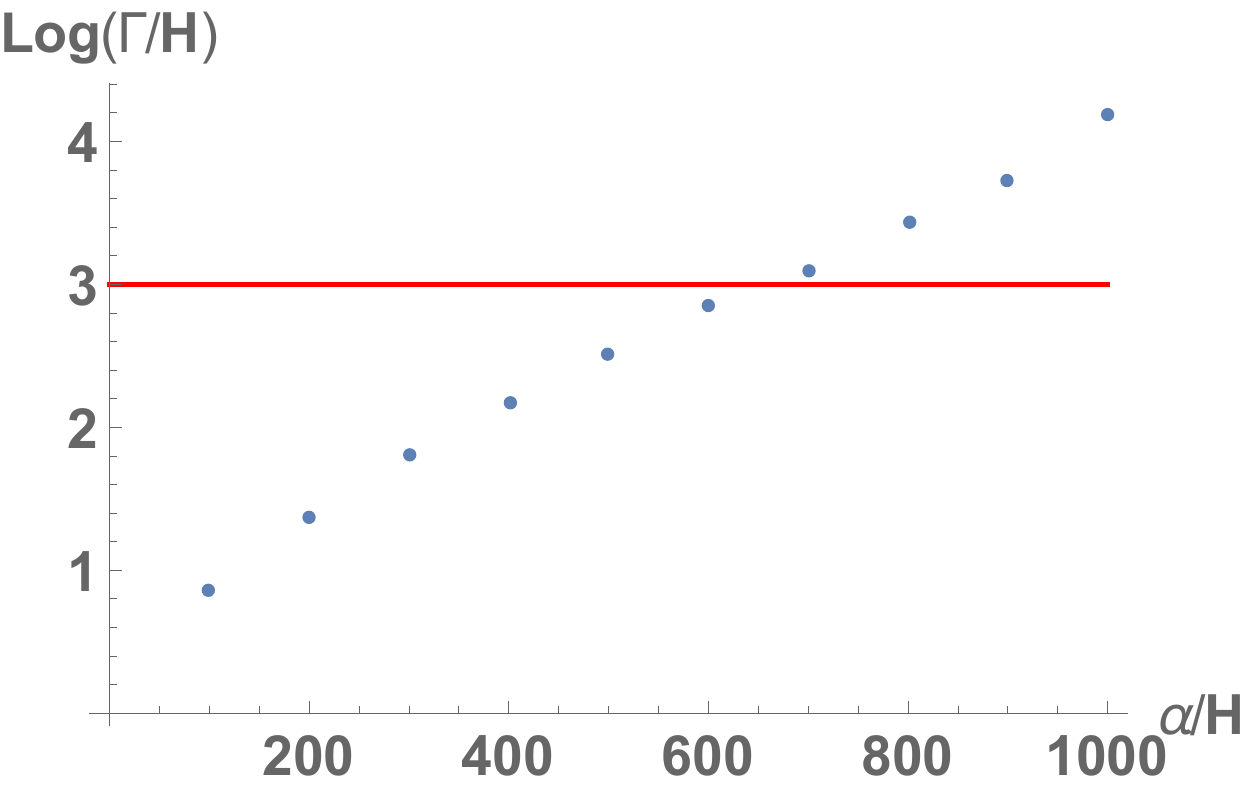}
\caption{(Left) The probability of remaining on the inflationary pleateau for a hilltop model (shown here for a cubic, $p=3$) decreases exponentially with time. (Right) The decay rate scales linearly with the parameter $\alpha$. For $\alpha > \alpha_{c,p}$, it grows larger than the Hubble expansion rate $3 H$ (shown in red), and eternal inflation shuts off. In the case of a cubic model shown here, $\alpha_{c,3} \approx 670 H$. }
\label{fig:fig}
\end{center}
\end{figure}

As shown in Figure \ref{fig:fig}, the probability of remaining on the inflationary plateau of a hilltop model ($V = V_0 - \frac{1}{p} \alpha \phi^p$, $p \geq 3$) decays exponentially in time. The decay rate $\Gamma$ grows roughly linearly with $\alpha$. For $\alpha$ larger than some critical $\alpha_{c,p}$, the decay rate is larger than the expansion rate $\exp 3 H t$, and inflation ceases to be eternal. The critical values discovered by our analysis for $p \leq 6$ are given roughly as follows:
\begin{equation}
\alpha_{c,p} \approx \left\{   \begin{array}{ccc}
\frac{3 \sqrt{6}}{2 \pi} H^3&p=1 &\text{(analytic)} \\
9 H^2 &p=2& \text{(analytic)}  \\
670 H&p=3  \\
1400 & p = 4 \\
3.4 \times 10^5H^{-1} & p = 5\\
1.8 \times 10^5H^{-2} & p = 6
\end{array} \right.
\label{eq:crit}
\end{equation}
Note that the $p=1$, $p=2$ cases in this table are the exact analytic expressions worked out in the previous subsection, whereas the $p \geq 3$ cases were computed numerically.

These results may be compared with the analytic estimates of \cite{Barenboim:2016mmw}. The analysis of that paper computed the characteristic time for fluctuations of the to grow to a critical value $\phi_c$, defined as the point at which the linear bound (\ref{eq:lineareternal}) is violated. They then treated this characteristic time as a lifetime for the field to exit the hilltop and demanded that this lifetime be shorter than the Hubble expansion time, which lead them to an analytic estimate of $\alpha_{c,p}$ of the form
\begin{equation}
\alpha_{c,p} H^{p-4}= \sqrt{3} \left( \frac{2 \pi}{\sqrt{3}} \right)^{p-2}.
\end{equation}
For $p =3,4,5,6$, this gives values of $2 \pi, 23, 83, 300$, respectively. These values exhibit a similar growth with $p$ that we observed in (\ref{eq:crit}), but they are orders of magnitude smaller than the values we found. This is in contrast with the linear and hilltop cases, in which our results agreed up to $O(1)$ factors with those derived in \cite{Barenboim:2016mmw}. The discrepancy in this general case is likely due to the fact that small deviations of a scalar field from the maximum of $p > 2$ hilltop, in contrast with a quadratic hilltop, do not grow exponentially with time. As a result, the field is able to ``hang on'' longer than one might have expected, and it takes a steeper slope to avoid eternal inflation.

Using $H^3 \Mp^2 = V/3$, our bounds on $\alpha$ translate to bounds on derivatives of the potential similar to (\ref{eq:lineareternal}) and (\ref{eq:hilltopeternal}). In order to support eternal inflation, the potential must satisfy
\begin{equation}
\left[ - \sgn(\partial^p V) \right]^{p+1} \frac{|\partial^p V|}{V^{(4-p)/2}} < (p-1)! \left( \frac{\alpha_{c,p}}{H^{4-p}} \right) (3 \Mp^2)^{(p-4)/2},
\label{eq:generalpeternal}
\end{equation}
where $\partial^p V$ is the $p$th derivative of $V$ and $\alpha_{c,p}/H^{4-p}$ is the numerical coefficient that appears in (\ref{eq:crit}).

\subsection{de Sitter Minima}\label{ssec:additional}

There is one other possible obstacle to eternal inflation. So far, we have focused on perturbative, tachyonic instabilities for de Sitter critical points, but one might also worry about non-perturbative instabilities, which affect even de Sitter minima of the potential. Semiclassical instabilities of this type require an exponentially-long lifetime \cite{Coleman:1980aw}, but this could still suffice to spoil eternal inflation if the Hubble expansion is also exponentially small, so that
\begin{equation}
\frac{\Gamma}{H^4} > \frac{9}{4\pi},
\end{equation}
with $\Gamma$ the decay rate per unit volume \cite{ArkaniHamed:2008ym}. In our universe $H \sim 10^{-60} \Mp$, so it is not crazy to imagine that this inequality could be satisfied for exponentially small $\Gamma/\Mp^4$. Indeed, as explored in \cite{ArkaniHamed:2008ym} (based on previous work of \cite{Isidori:2001bm, Coleman:1977py}), even a slight modification of the parameters of the Standard Model could lead to a vacuum decay rate that is larger than the Hubble constant, resulting in a vacuum that would decay before it could eternally inflate. Based on the subsequently-measured value of the Higgs mass, it appears that the Standard Model vacuum is sufficiently long-lived to yield eternal inflation, but as we will discuss in \S\ref{sec:SWAMPLAND}, this result is sensitive to UV physics.


\section{Eternal Primordial Inflation}\label{sec:PRIMORDIAL}

So far, we have derived conditions for eternal inflation at general points of the scalar potential. In this section, we apply these criteria to a handful of popular models of primordial inflation, to see if/when they become eternal. As a warm-up, we consider power-law models of (chaotic) inflation, in which the conditions for eternal inflation are well-known, before moving on to Starobinsky inflation, hilltop inflation, and inflection point inflation.

\subsection{Power-Law Inflation}

Power-law (chaotic) inflation \cite{Linde:1983gd} features a potential of the form
\begin{equation}
V(\phi) = \alpha \phi^p.
\end{equation}
Inflation happens for large $\phi$, and it ends when the field rolls to its minimum at $\phi = 0$. To get around 60 $e$-folds of inflation, the inflaton must start at a pivot scale of about $\phi_* \approx \sqrt{120 p} \Mp$. The power spectrum amplitude is given by 
\begin{equation}
A_s \approx \frac{1}{24 \pi^2} \frac{1}{\Mp^4} \frac{ V_* }{ \epsilon_{V,*} } \approx 2 \times 10^{-9},
\label{eq:As}
\end{equation}
where $*$ indicates evaluation at $\phi=\phi_*$, and the first slow-roll parameter is given by
\begin{equation}
\epsilon_V = \frac{\Mp^2}{2} \left( \frac{V'(\phi)}{V(\phi}\right)^2 = \frac{\Mp^2}{2} \frac{p^2}{\phi^2}.
\end{equation}
Plugging this into the previous equation, we find
\begin{equation}
\alpha = \frac{2 \pi^2 p \times 10^{-10}}{ (120 p)^{p/2} } \Mp^{4-p}.
\end{equation}
This means
\begin{equation}
\frac{V'(\phi)}{V(\phi)^{3/2}} = \frac{1}{\pi \sqrt{2}} p^{(p+2)/4} \times 120^{p/4} \times 10^5 \times \phi^{-(p+2)/2} \Mp^{(p-4)/2}.
\end{equation}
By (\ref{eq:lineareternal}), inflation becomes eternal when $V'/V^{3/2} < \sqrt{2}/(2 \pi \Mp^3)$. Solving for $\phi$, we find
\begin{equation}
\phi_{\rm{EI}} =p^{1/2} 2^{\frac{p}{p+2}} \times 3^{\frac{p}{2(p+2)}}\times 10^{\frac{p+20}{2(p+2)}} \Mp \approx \Mp \times \left\{\begin{array}{cc}
11000 & p =\frac{1}{2} \\
8000 & p =\frac{2}{3} \\
5000  & p =1\\
1500 & p =2\\
700 & p =3\\
500  & p =4\\
\end{array} \right.
\label{eq:eternalchaotic}
\end{equation}
For $p \gtrsim 4$, $V(\phi_{\rm{EI}}) \gtrsim \Mp^4$, indicating that these models break down before $\phi$ gets large enough for eternal inflation to occur. But even for models with $p \leq 4$, we see that $\phi_{\rm EI} \gg \phi_* \approx \sqrt{120p} \Mp$. If higher-order terms in the potential (presumably due quantum gravity effects) become relevant for $\phi_* < \phi < \phi_{\rm EI}$, it is easy to imagine that the model could give 60 $e$-folds of inflation, yet fail to be eternal.

Of course, these power-law models have other issues: even the modestly super-Planckian traversal $\Delta \phi = \phi_* = \sqrt{120p} \Mp$ is at odds with effective field theory, which suggests that Planck-suppressed operators of the form $\phi^n/\Mp^{n-4}$ should spoil the flatness of the inflationary potential for $\phi$ larger than $\Mp$. More importantly, the latest \emph{Planck} constraints on the spectral index $n_s$ and tensor-to-scalar ratio $r$ rule out power-law models with $p > 2$ at more than $95 \%$ confidence, and power-law models with $p < 1$ do not seem to be the most likely candidates either \cite{Akrami:2018odb}. Thus, we are led to consider models in better agreement with observation.

\subsection{Starobinsky Inflation}

One such model is Starobinsky inflation \cite{Starobinsky:1980te}. The potential is given by
\begin{equation}
V(\phi) = V = V_0 \left(1 - \exp \left[ -\sqrt{2/3} \,\phi/\Mp \right] \right)^2,
\end{equation}
Inflation occurs for positive $\phi$, and ends when the field rolls to the minimum at $\phi = 0$. predicts $r \approx 0.004$, $n_s \approx 0.963$, in excellent agreement with observation. To get around 60 $e$-folds of inflation, the field must begin at a value of $\phi_* \approx 5.5 \Mp$. Setting $A_s = 2 \times 10^{-9}$ and using (\ref{eq:As}) gives $V_0 \approx 8 \times 10^{-11} \Mp^4$. This in turn means that eternal inflation sets in for
\begin{equation}
\phi_{\rm EI} \approx 16 \Mp,
\end{equation}
when (\ref{eq:lineareternal}) is satisfied. Due to the exponential flatness potential, we see that eternal inflation sets in much closer to $\phi_*$ than in the power-law inflation models. If one can find a way to protect the Starobinsky model from Planck-suppressed operators for $0 < \phi < \phi_*$, it may not be too much harder to protect the model over the range $0 < \phi < \phi_{\rm EI}$, thereby achieving eternal inflation. However, as in the power-law inflation case, such super-Planckian traversals are likely difficult to achieve once the effects of quantum gravity are considered. As a result, we turn our attention to models that do not necessarily require such super-Planckian excursions of the inflaton.

\subsection{Quadratic Hilltop Inflation}

We consider quadratic hilltop models of inflation with a potential of the form
\begin{equation}
V = V_0 - \frac{1}{2} m^2 \phi^2 + \frac{1}{6} \lambda \phi^3 + ...
\label{eq:hilltoppot}
\end{equation}
Inflation occurs near the hilltop at $\phi = 0$. The phenomenology of these models has been studied in e.g. \cite{Linde:2007jn}, and they arise with regularity in models of random inflation \cite{Masoumi:2016eag, Rudelius:2018yqi}. These models can be either small-field $(\phi < \Mp)$ or large-field ($\phi > \Mp$). In the small-field case, the spectral index is given by \cite{Linde:2007jn}
\begin{equation}
n_s \approx 1 - 2 \frac{m^2 \Mp^2}{V_0} \coth \left(\frac{m^2 \Mp^2 N_e}{2 V_0} \right),
\label{eq:hillspec}
\end{equation}
with $N_e \lesssim 60$. This achieves a maximum of $n_s = 0.933$ as $m^2 \Mp^2/V_0 \rightarrow 0$, which is too small to agree with observation. Large-field quadratic hilltop models do not seem to be the most likely candidates for explaining the data, as they typically tend to favor smaller values of $n_s$, but they are not incompatible with the $n_s \approx 0.965$ value favored by \emph{Planck} \cite{Rudelius:2018yqi}. A special case of large-field quadratic hilltop models are models of natural inflation, which have
\begin{equation}
V(\phi) = \frac{V_0}{2} \left(1 + \cos \frac{\phi}{f} \right) = V_0 - \frac{V_0}{4} \frac{\phi^2}{f^2} + ...,
\label{eq:naturalexpansion}
\end{equation}
i.e. they can be thought of as quadratic hilltop models with $m^2 = V_0/2f^2$.

Let us consider the prospects of eternal inflation in these models. The first derivative of the potential vanishes when $\phi = 0$, thereby satisfying (\ref{eq:lineareternal}), so the key question is whether or not (\ref{eq:hilltopeternal}) is satisfied. This occurs when
\begin{equation}
\eta_V = \frac{m^2 \Mp^2}{V_0} < 3,
\end{equation}
where $\eta_V := \Mp^2 V''(\phi)/V$ is the second slow-roll parameter. Phenomenologically-viable models require $\eta_V \ll 1$, so this constraint is always satisfied, and inflation is necessarily eternal.

\subsection{Inflection Point Inflation}

Next, we consider models of inflection point inflation, which have a potential of the form
\begin{equation}
V = V_0 + \alpha \phi + \frac{1}{6} \lambda \phi^3 + ....
 \label{eq:inflectionpot}
\end{equation}
Inflation occurs near the inflection point at $\phi = 0$. The phenomenology of these models was considered in \cite{Linde:2007jn, Baumann:2007ah}, and they arise with regularity in models of random inflation \cite{Masoumi:2016eag, Rudelius:2018yqi} as well as string theory models of D-brane inflation on the conifold \cite{Agarwal:2011wm}. Like with quadratic hilltop models, inflection point models can be either small-field or large-field. But unlike quadratic hilltop inflation, the spectral index in the small-field case is actually in good agreement with observation. The spectral index given by \cite{Linde:2007jn}
\begin{equation}
n_s \approx 1 - \frac{4 \pi}{\Ntot} \cot \left( \frac{\pi N_e}{\Ntot} \right),
\label{eq:inflecspec}
\end{equation}
where $N_e \lesssim 60$ and $\Ntot$ is the total number of $e$-folds, which is given by 
\begin{equation}
\Ntot \approx \pi \sqrt{2}  \frac{V_0}{\sqrt{\alpha \lambda}}.
\end{equation}
The range of this function is complementary to the one for quadratic hilltop inflation (\ref{eq:hillspec}): here, $n_s > 0.933$ for $N_e = 60$, with $n_s$ approaching the lower bound $0.933$ as $\Ntot \rightarrow \infty$ and growing larger as $\Ntot \rightarrow 60$. Agreement with experiment occurs for $120 \lesssim \Ntot \lesssim 200$, allowing $N_e$ to vary between 50 and 60.

Phenomenologically-viable, small-field models of inflection point inflation are not eternal. This follows from equation (\ref{eq:As}):
\begin{equation}
 \frac{(V'(\phi_*))^2 }{  V_*^3 } \approx  \frac{5 \times 10^{10}}{24 \pi^2} \frac{1}{\Mp^6},
\end{equation}
with $\phi_*$ close to the inflection point. This clearly violates (\ref{eq:lineareternal}), which means that the potential is not sufficiently flat at the inflection point to generate eternal inflation.

On the other hand, large-field inflection point models can be eternal provided that $\phi_*$, the field value 60 $e$-folds before the end of inflation, is a sufficiently-large distance away from the inflection point, $\phi_{\rm inf}$. In this situation, the potential might be much flatter at the inflection point than it is at $\phi_*$, so (\ref{eq:lineareternal}) is obeyed at the former while (\ref{eq:As}) is satisfied at the latter.

As a concrete example, consider an inflection point model (\ref{eq:inflectionpot}) with $V_0 = 2.8 \times 10^{-10} \Mp^4$, $\alpha = 1.6 \times 10^{-26} \Mp^3$, $\lambda = 1.0 \times 10^{-12} \Mp$. This gives $r_* \approx .009$, $n_{s,*} \approx 0.964$, $A_s \approx  2 \times 10^{-9}$, which is in good agreement with observation. Note that a na\"ive application of (\ref{eq:inflecspec}) would indicate $n_{s,*} = 0.933$, which is not correct. This discrepancy is due to the fact that $\phi_*$ is located a super-Planckian distance away from the inflection point (to be precise, $\phi_* - \phi_{\rm inf} \approx 4.3 M_p$), and the field rolls a distance of larger than $5 M_p$ during its last 60 $e$-folds. The approximation used to compute (\ref{eq:inflecspec}), namely, that the potential $V(\phi)$ is roughly constant during slow-roll, is not valid in this large-field context. As a result, we have $V'(\phi_{\rm inf})/V^{3/2}(\phi_{\rm inf}) \approx 3 \times 10^{-10} \Mp^{-3}$ and $V'''(\phi_{\rm inf})/V^{1/2}(\phi_{\rm inf}) \approx 6 \times 10^{-8} \Mp^{-1}$,  so (\ref{eq:lineareternal}) and (\ref{eq:generalpeternal}) are both satisfied by many orders of magnitude, and eternal inflation occurs at the inflection point. Similar examples of eternal inflation in inflection point models can be found in \cite{Hertog:2015zwh}.

\subsection{General Hilltop Inflation}

Finally, we consider a more general hilltop potential of the form
\begin{equation}
V = V_0 - \frac{1}{p} \alpha \phi^p\,,~~~p \geq 4.
\end{equation}
From the outset, we should note that these models seem very fine-tuned from the perspective of effective field theory, since a degree $p$ hilltop requires all of the $\phi^q$ series coefficients to vanish for $q \leq p$ at the origin. For this reason, we tend to think that such hilltops will be much rarer than inflection points and quadratic hilltops in the string Landscape. Nonetheless, we proceed with our analysis of phenomenology and eternal inflation in these models.

We saw in \S\ref{ssec:generalhill} that these models will be eternal unless $\alpha$ is very large in Hubble units, which in turn implies $\alpha/V_0 \gg 1/\Mp^p$. Working in this limit allows us to make some simplifying approximations in our analysis. To begin, we may write
\begin{equation}
\epsilon_V(\phi) \approx \frac{\Mp^2}{2} \left( \frac{  \alpha \phi^{p-1} }{ V_0  }  \right)^2 \,,~~~  \eta_V(\phi) \approx -\frac{(p-1)\alpha \Mp^2}{V_0}\phi^{p-2.}
\label{eq:hillform}
\end{equation}
One can check that $\epsilon_V(\phi) > 1$ for $\phi = \Mp$, so large-field hilltop models are necessarily eternal. The pivot scale $\phi_*$ $N_e$ $e$-folds before the end of inflation is given by
\begin{equation}
\phi_* = \left[\alpha (p-2) N_e \frac{ \Mp^2}{V_0} \right]^{-1/(p-2)}.
\label{eq:hillphistar}
\end{equation}
In the limit of interest, $\epsilon_V(\phi_*)$ is negligibly small, which means that the tensor-to-scalar ratio will be unobservable. The running of the spectral index $\alpha_s$ is given by
\begin{equation}
\alpha_s \approx -\frac{(p-1)}{(p-2)}\frac{2}{N_e^2},
\end{equation}
which is also small and in agreement with observation. On the other hand, the spectral index $n_s$ itself is given by
\begin{equation}
n_s \approx 1 - \frac{(p-1)}{(p-2)} \frac{2}{N_e}. 
\end{equation}
For $N_e =60$, this gives $0.95$, $0.956$, $0.958$ for $p=4,5,6$, respectively, and asymptotes to $0.967$ as $p \rightarrow \infty$. Comparing to the 2018 \emph{Planck} constraints of $n_s = 0.9649 \pm 0.0042$ at $68 \%$ CL \cite{Akrami:2018ylq}, we see that the $p=4$ case is ruled out at more than $3 \sigma$, the $p=5$ case is on the verge of being excluded at $2\sigma$, but the $p \geq 6$ cases are all in good agreement with observation.

So far, there does not seem to be anything wrong with a non-eternal hilltop model. However, we must further impose the constraint $A_s \approx 2 \times 10^{-9}$. Plugging in our expressions (\ref{eq:hillform}) and (\ref{eq:hillphistar}) for $\epsilon_V(\phi_*)$ into the formula for $A_s$ in (\ref{eq:As}), we find
\begin{equation}
\alpha H^{p-4} = \frac{(12 \pi^2 \times 2 \times 10^{-9})^{(p-2)/2}}{3^{(p-4)/2} \left[(p-2) N_e\right]^{p-1}} \ll 1,
\end{equation}
which implies by (\ref{eq:crit}) that these models are necessarily eternal. This agrees with the result of \cite{Barenboim:2016mmw}.


\section{Eternal Inflation and the Swampland}\label{sec:SWAMPLAND}

The search for controlled de Sitter solutions in string theory has been going on for many years, but recently the subject has received renewed attention, due in large part to a conjectured bound on potentials in string theory, which would forbid metastable de Sitter vacua altogether. The (refined) de Sitter Conjecture (RdSC) of \cite{Garg:2018reu, Ooguri:2018wrx,Obied:2018sgi} holds that scalar potentials in a consistent theory of quantum gravity must satisfy one of the following two bounds at every point in field space:
\vspace{.2 cm}
\begin{equation}
| \nabla V| \geq c  \cdot V,
\label{eq:dSC}
\end{equation}
OR
\begin{equation}
\text{min} (\nabla_i \nabla_j V) \leq -c' \cdot V,
\label{eq:RdSC}
\end{equation}
for some positive universal constants $c, c' \sim O(1)$ in Planck units. The left-hand side of (\ref{eq:RdSC}) should be understood as the smallest eigenvalue of the Hessian in an orthonormal frame. Note that this conjecture is trivially satisfied if $V \leq 0$, since the left-hand side of (\ref{eq:dSC}) is manifestly non-negative.

A closely-related conjecture of this sort, which we will call the RdSC$^*$, has been put forward in \cite{Andriot:2018mav}. It holds that in any point in field space with $V>0$,
\begin{equation}
\left( \Mp \frac{|\nabla V|}{V} \right)^{q} - a \Mp^2 \frac{\text{min} \nabla_i \nabla_j V}{V} \geq b\,,~~~~~~~\text{with } a+b=1,~~a, b > 0,~~q> 2,
\label{eq:RdSC2}
\end{equation}
for some constants $a, b$, and $q$, which have yet to be precisely determined.

As of yet, neither of these conjectures have accumulated convincing evidence. Both of them have been verified to hold for suitable values of the constants $c, c', a,b, q$ in a number of well-controlled examples in string theory, but more sophisticated constructions of de Sitter vacua, notably the KKLT scenario \cite{Kachru:2003aw} and the LVS scenario \cite{Conlon:2005ki}) have also been claimed, which (if valid) would represent counterexamples to these conjectures. Debates over the validity of these conjectures and the claimed de Sitter constructions are currently in progress, and we do not attempt to resolve them here.

Instead, we wish to draw attention to the remarkable similarities between the bounds hypothesized in the RdSC and the RdSC$^*$ and the conditions we have derived on eternal inflation. In particular, we found that to evade eternal inflation, at least one of the following must be true:
\begin{enumerate}
 \item The linear bound (\ref{eq:lineareternal}) is violated, so that
\begin{equation}
\frac{|\nabla V|}{V} > \frac{\sqrt{2 V}}{2 \pi \Mp^3} 
\label{eq:nolinear}
\end{equation} 
\item The quadratic bound (\ref{eq:multihilltopeternal}) is violated, so that
\begin{equation}
\frac{\sum_{i} \nabla_i \nabla_i V(\bm{\phi})}{V(\bm{\phi})} < -\frac{3}{M_{\rm Pl}^2},
\label{eq:nohilltop}
\end{equation}
with the sum here taken over just the negative eigenvalues of the Hessian.
\item The combined bound (\ref{eq:combined}) is violated, so that
\begin{equation}
\frac{2 \pi^2 \Mp^2 |\nabla V|^2}{ V^2} - \frac{V}{3 \Mp^2} \frac{\sum_{i}  \nabla_i \nabla_i V}{V} > \frac{V}{\Mp^4} ,
\label{eq:nocombined}
\end{equation}
with the sum over $i$ again running over the negative eigenvalues of the Hessian.
\item Some higher-derivative of the potential is sufficiently large, so that (\ref{eq:generalpeternal}) is violated.
Assuming that the dynamics are effectively single-field, the condition for avoiding eternal inflation is given by
\begin{equation}
\left[ - \sgn(\nabla^p V) \right]^{p+1} \frac{|\nabla^p V|}{V^{(4-p)/2}} > \mathcal{N}_p \Mp^{p-4},
\label{eq:higherderiv}
\end{equation}
with 
\begin{equation}
\mathcal{N}_p \approx \left\{   \begin{array}{cc}
770 &p=3  \\
8400 & p = 4 \\
1.4 \times 10^7 & p = 5\\
6.5 \times 10^7& p = 6 \\
... & ...
\end{array} \right..
\end{equation}

For a sum-separable potential $V(\bm{\phi}) = \sum_i V_i(\phi^i)$, we expect that the decay rates will add linearly, so (\ref{eq:nocombined}) generalizes to
\begin{equation}
\left(\frac{V}{3 \Mp^6}\right)^{1/2} \left( \sum_{i=1}^N \Gamma_i(\phi^i) \right) > \frac{V}{\Mp^4},
\end{equation}
with $\Gamma_i(\phi^i)$ dictated by the lowest non-negligible derivative of $V_i(\phi^i)$,
\begin{equation}
\Gamma_i(\phi^i) \in \left\{ \frac{2 \pi^2\sqrt{3} \Mp^5|\nabla_i V|^2}{V^{5/2}}, - \frac{\Mp \nabla_i^2 V}{\sqrt{3V}} ,...,\left[ - \sgn(\nabla^p V) \right]^{p+1} \frac{\sqrt{3} |\nabla_i^p V| V^{(p-3)/2}}{\mathcal{N}_p\Mp^{p-3}} ,...\right\},
\end{equation}
where we set $\Gamma_i =0$ if the above expression is negative (which will occur if the $p$th derivative is positive and $p$ is even). 

\item There is a non-perturbative instability (as discussed in \S\ref{ssec:additional}) with decay rate per unit volume $\Gamma$ satisfying
\begin{equation}
\frac{\Gamma}{H^4} > \frac{9}{4\pi}.
\label{eq:nodecay}
\end{equation}
\end{enumerate}
Comparing the bounds (\ref{eq:dSC}) to (\ref{eq:nolinear}),  (\ref{eq:RdSC}) to (\ref{eq:nohilltop}), and (\ref{eq:RdSC2}) to (\ref{eq:nocombined}), the similarities are striking.\footnote{The higher-derivative bound in (\ref{eq:higherderiv}) is likewise reminiscent of the conditions needed to avoid ``generalized'' slow-roll described in \cite{Garg:2018zdg}.}

Indeed, has previously been noted that for suitable values of the constants $c$, $c'$ the RdSC bounds (\ref{eq:dSC}) and (\ref{eq:RdSC}) are incompatible with the bounds (\ref{eq:nolinear}) and (\ref{eq:nohilltop}), so eternal inflation is incompatible with the RdSC \cite{Matsui:2018bsy, Dimopoulos:2018upl, Kinney:2018kew, Brahma:2019iyy}. To see this, we simply use the fact that $V < \Mp^4$ and the fact that $|\sum_i \nabla_i \nabla_i V| \geq |\text{min} \nabla_i \nabla_j V|$, in which case (\ref{eq:dSC}) implies (\ref{eq:nolinear}) provided $c > \sqrt{2}/2\pi \Mp^{-1}$, and (\ref{eq:RdSC}) implies (\ref{eq:nohilltop}) provided $c' > 3 \Mp^{-2}$.

It is also worth pointing out that the RdSC$^*$ bound (\ref{eq:RdSC2}) for $q = 2$ and suitable values of $a$ and $b$ implies the bound (\ref{eq:nocombined}). To see this, we multiply both sides of the RdSC$^*$ bound by $2 \pi^2 V/\Mp^4$, then set $V < \Mp^4$ and $|\sum_i \nabla_i \nabla_i V| > |\text{min} \nabla_i \nabla_j V|$ to get
\begin{equation}
2 \pi^2 \left( \Mp \frac{|\nabla V|}{V} \right)^{2} - 2 \pi^2 a \frac{V}{\Mp^2} \frac{\sum_i \nabla_i \nabla_i V}{V} \geq 2 \pi^2 b \frac{V}{\Mp^4}.
\end{equation}
This implies (\ref{eq:nocombined}) provided $2 \pi^2 a > 1/3$, $2 \pi^2 b < 1$, which is indeed consistent with $a+b=1$.

There are a few possible conclusions that could be drawn from the agreement between these recently-proposed Swampland criteria with the conditions for evading eternal inflation. One is that it is simply a coincidence, or perhaps a lamppost effect: the RdSC bounds hold only in weakly coupled regimes in string theory, and more sophisticated string theory constructions will violate all of these bounds, enabling both de Sitter vacua and eternal inflation. Even this possibility is interesting, however: if the RdSC holds in weakly coupled string theory, then by our analysis above, stochastic eternal inflation is forbidden in such regimes, which means that eternal inflation can only take place in strongly coupled string theory. This coheres well with the picture of \cite{Hawking:2017wrd}, which argued that eternal inflation must be understood within a genuinely quantum gravitational regime.

A second possibility, which has been adopted by several previous works on the subject, is that there may be some deep, fundamental principle of quantum gravity that explains why these Swampland criteria must hold, which in turn forbids eternal inflation as corollary.\footnote{One candidate ``fundamental principle'' was in fact proposed in \cite{Ooguri:2018wrx}, relating the RdSC to the Swampland Distance Conjecture \cite{Ooguri:2006in} and the Covariant Entropy Bound \cite{Bousso:1999xy}, but it is unclear that this argument should apply outside of parametrically-controlled regions in string theory, where de Sitter vacua have long been presumed absent \cite{Dine:1985he}.} 

But a third possibility, which has not received significant attention, is that the condition of No Eternal Inflation is \emph{itself} the deep, fundamental principle that explains why these Swampland criteria should (approximately) hold true.\footnote{See however \cite{Dvali:2018jhn}, which noted similarities between the RdSC bounds and the bounds required to avoid de Sitter ``quantum breaking'' \cite{Dvali:2013eja}, which would in turn forbid eternal inflation.} In other words, rather than viewing the RdSC and RdSC$^*$ as the more fundamental condition from which No Eternal Inflation follows as a consequence, perhaps we should view the No Eternal Inflation condition as the more fundamental condition from which the RdSC and RdSC$^*$ approximately emerge. One might view this as a lateral move, since there is not an obvious, deep reason why eternal inflation should be incompatible with quantum gravity, either. However, eternal inflation is the \emph{type} of thing that quantum gravity \emph{might} eschew, as it leads to a qualitatively different sort of universe than a theory without eternal inflation, whereas there does not seem to be anything qualitatively different between theories that narrowly violate the RdSC bound (\ref{eq:dSC}) and those that narrowly satisfy it. Additionally, the No Eternal Inflation bounds are slightly weaker than the RdSC$^{(*)}$ bounds, which means that (a) the evidence presented for the RdSC$^{(*)}$ thus far in the form of string theory examples and no-go theorems also serves as evidence for a No Eternal Inflation principle, yet (b) the constraints from the RdSC$^{(*)}$ on phenomenology and string theory model-building are relaxed in a favorable way for a No Eternal Inflation principle, as we will see shortly.

With this motivation in mind, we devote the remainder of this paper to speculating about the consequences of a No Eternal Inflation principle. We stress that we do not by any means claim this principle to be a theorem, nor do we even consider it to be a well-supported conjecture, but rather we view it as a ``supposal'': supposing that eternal inflation is, for some yet unknown reason, incompatible with quantum gravity: what consequences would this have for scalar potentials in string theory, and phenomenology in our own universe? This possibility is certainly worth pondering, and we are not the first to do so (see e.g. \cite{ArkaniHamed:2008ym, Banks:2003pt, Page:2006nt, Dvali:2013eja}), but in light of the rekindled debate on de Sitter vacua in string theory, it seems like a good time to consider the question afresh. However, we must take care to acknowledge the distinct possibility that we are chasing the rabbit down the wrong hole. Many readers may prefer to retreat to the first possibility: eternal inflation is perfectly compatible with quantum gravity, and the de Sitter conjectures are simply consequences of a lamppost effect. But for those who are up for the adventure, we pose the question: what if eternal inflation is in the Swampland?

\subsection{Quintessence vs. Cosmological Constant}

Aside from the lack of a compelling physical motivation, probably the biggest issue facing the RdSC$^{(*)}$ is the observed dark energy in our universe, coupled with the fact that realistic quintessence models seem to be at least as difficult to construct in string theory as metastable de Sitter vacua \cite{Hellerman:2001yi, Cicoli:2018kdo, Hertzberg:2018suv}. Furthermore, even hypothetical quintessence models that satisfy the RdSC bounds are on the verge of being ruled out experimentally \cite{Agrawal:2018own, Heisenberg:2018yae, Colgain:2019joh, Akrami:2018ylq}.

By contrast, the bounds obtained by forbidding eternal inflation are modest enough to resolve these issues. For a quintessence model, replacing the lower bound on $V'/V$ in (\ref{eq:dSC}) with a lower bound on $V'/V^{3/2}$ in (\ref{eq:nolinear}), we see that the tension between the observed dark energy and the RdSC with $c \gtrsim \Mp^{-1}$ can be alleviated: potentials of the form $V(\phi) \sim e^{-\lambda \phi}$ for $\phi$ large can have $\lambda \ll \Mp^{-1}$ and still satisfy the modified bound (\ref{eq:nolinear}).\footnote{Of course, this is not necessarily a good thing for fans of the RdSC, since it would make the conjecture very difficult to test experimentally: one of the advantages of the original bound (\ref{eq:dSC}) is that it makes predictions which could be tested in the foreseeable future.}

But an even more welcome possibility is a metastable de Sitter vacuum with a decay rate per unit volume $\Gamma$ that satisfies (\ref{eq:nodecay}). This possibility is of course not allowed by the RdSC and RdSC$^*$, but it is compatible with a No Eternal Inflation principle, since the vacuum decays before it eternally inflates. Since $\Gamma/\Mp^4$ is necessarily exponentially-suppressed for a semiclassical instability, this inequality implies that metastable de Sitter vacua which decay via a semiclassical instability must have exponentially-small Hubble rate. At best, this could be viewed as a partial explanation for the smallness of the cosmological constant in our universe, $\Lambda \sim H^2 \Mp^2 \sim 10^{-120} \Mp^4$. At least, it is worth noting that the observed cosmological constant in our universe is small enough to allow for a semiclassical instability consistent with (\ref{eq:nodecay}).

In \cite{ArkaniHamed:2008ym}, a 2-loop running computation of the Higgs potential examined the prospects of eternal inflation in the Standard Model vacuum. Assuming all parameters of the Standard Model except the Higgs mass are held fixed, the authors found that the decay rate would be large enough to prevent eternal inflation for $m_H \lesssim 115$ GeV, whereas for $m_H \gtrsim 130$ GeV, the decay channel disappears entirely, and for $m_H \lesssim 110$ GeV, the decay rate is too large to agree with observation (which requires the lifetime of the universe to be $O(H^{-1})$). The subsequent measurement of $m_H \approx 125$ GeV \cite{Aad:2012tfa, Chatrchyan:2012xdj, Aad:2015zhl} indicates that the Standard Model vacuum is sufficiently long-lived to produce eternal inflation \cite{Degrassi:2012ry}.

This might seem to rule out any prospects of a No Eternal Inflation principle, but such a conclusion is premature for a couple of reasons. First of all, as noted in \cite{ArkaniHamed:2008ym}, one expects that our vacuum will admit alternative semiclassical decay channels into other vacua in the Landscape, one or more of which might have a lifetime that is short enough to prevent eternal inflation. Furthermore, one can check that the Standard Model vacuum decay rate is exponentially-sensitive to UV physics at scales well above $10^6$ GeV. A heavy fermion doublet with a sufficiently-large Yukawa coupling to the Higgs, for instance, could conceivably lower the decay rate $\Gamma$ so that it satisfies (\ref{eq:nodecay}), thus inhibiting eternal inflation.\footnote{Conversely, in a sufficiently high-scale model of primordial inflation, bubbles with large Higgs vev can quickly come to dominate the universe \cite{Hook:2014uia, Kearney:2015vba, East:2016anr}. This suggests that UV physics should actually stabilize the Standard Model vacuum during primordial inflation up to energies of order the inflationary Hubble scale, which could make it more difficult to avoid eternal inflation via a Higgs tunneling instability. We plan to investigate this possibility in more detail in the near future} Finally, one could imagine that vacuum decay occurs more rapidly through some non-semiclassical process that depends on unknown Planck-scale physics.

 One may object that these possibilities seem to require a great deal of fine-tuning. After all, the fact that we still exist after a Hubble time $H^{-1}$ indicates that the decay rate per unit volume $\Gamma$ of our universe cannot be much larger than $9 H^4/4\pi$, so it requires a very precise tuning of $\Gamma$ to avoid eternal inflation without destroying our universe. But one could also turn this argument around: if inflation \emph{is} eternal, and the lifetime of our universe is much larger than $H^{-1}$ (as suggested by the 2-loop Standard Model analysis \cite{Degrassi:2012ry}), the fact that we do exist so close to the Hubble time, famously known as the ``cosmic coincidence problem,'' is left unexplained. Other solutions to the cosmic coincidence problem have been proposed: for instance, certain cosmological measures proposed on the eternal inflation multiverse, such as the causal patch measure \cite{Bousso:2006ev}, seem to favor observers that live within a Hubble time after the end of inflation \cite{Bousso:2010zi}. However, the bound (\ref{eq:nodecay}) implied by a No Eternal Inflation principle would also ensure that observers in a metastable de Sitter vacuum cannot exist beyond a time of $O(H^{-1})$, and the fine-tuning required to ensure that our vacuum survives for at least a Hubble time could partially be explained anthropically \cite{Weinberg:1988cp}, since gravitational collapse leading to galaxy formation did not occur until a time $t = O(10^{-1}) H_0^{-1}$. Of course, this need not be true in general: one could imagine a universe with a cosmological constant that is orders of magnitude smaller than our own, in which case galaxy formation may happen well before the time of dark energy domination. But then again, this question persists even if one allows for eternal inflation, and its resolution is likely due to the distribution of (anthropically-viable) vacuum energies in quantum gravity, about which little is known.\footnote{See however \cite{Bousso:2000xa} for some preliminary work on this question.} 
 
A No Eternal Inflation principle, unlike the the RdSC$^{(*)}$, is consistent with claimed de Sitter constructions in string theory, with some caveats: to avoid eternal inflation, the decay rate of these vacua must be $O(H)$, which is much larger than the decay rate of KKLT vacua for tunneling through the potential barrier with $V(\phi) > 0$ to the large-volume region of parameter space \cite{Kachru:2003aw}. This implies that some other instability must arise, though for $H \ll \Mp$, this could be a semiclassical, non-perturturbative decay to another vacuum, as discussed previously in the context Standard Model. Possible instabilities in the KKLT construction have been discussed in the literature \cite{Bena:2014jaa, Bena:2015kia,  Bena:2016fqp, Danielsson:2015eqa}, though many of these challenges have been addressed \cite{Michel:2014lva, Kachru:2018aqn}.

Similarly, a No Eternal Inflation principle is consistent with the data from classical de Sitter supergravity solutions, with some caveats. In \cite{Andriot:2018mav}, a number of de Sitter critical points or near-critical points ($V'(\phi)/V \ll 1/\Mp$) were examined in the context of classical IIA supergravity. 19 of the 20 near-critical points presented had $\eta_V :=\Mp^2 \min_i (\nabla_i \nabla_i V)/V < -3$, indicating that they necessarily satisfy condition (\ref{eq:nohilltop}), forbidding eternal inflation. One near-critical point had $\eta_V \approx -2.5$, which would violate (\ref{eq:nohilltop}) and presumably lead to eternal inflation, assuming that the other negative eigenvalues of the Hessian are sufficiently close to 0. However, all of the classical dS solutions found in that paper required either small volume, large string coupling, or a large number of orientifold planes, rendering such solutions untrustworthy within the supergravity regime. Likewise, \cite{Flauger:2008ad, Caviezel:2008tf} found examples of near-critical points with $\eta_V \approx -2.5$, but these suffer from the same questions of reliability within the supergravity framework. In the type IIB supergravity context, \cite{Caviezel:2009tu} was unable to find any near-critical points with $\eta_V > -3.1$. It is mildly intriguing that de Sitter solutions in classical supergravity seem hard to find for $\eta_V$ much smaller than the critical value of $\eta_V = -3$ necessary for eternal inflation. It is rather exciting that the bound (\ref{eq:nohilltop}) needed to avoid eternal inflation is strong enough to come into contact with data from these classical supergravity solutions, so it may be possible to falsify a No Eternal Inflation condition in the near future by further constructions of de Sitter critical points in string theory. Indeed, several recent works may have already made progress along this direction \cite{Blaback:2018hdo, Blaback:2019zig, Terrisse:2019usq}, though to firmly establish one of these constructions as a counterexample, one would need to check that the truncated modes do not yield instabilities that spoil eternal inflation.

\subsection{Primordial Inflation}

What implications would a No Eternal Inflation principle have for inflationary observables? We saw in \S\ref{sec:PRIMORDIAL} that certain models (namely, power-law, Starobinsky, and inflection point) can be made non-eternal, whereas other models (namely, hilltop and natural) are necessarily eternal at the maximum of their potentials. Of these models, power-law, Starobinsky, and natural inflation models are necessarily large-field, while inflection point and hilltop models can be small-field.

For large-field models, protecting the potential from Planck-suppressed operators is a challenge for embedding the models in a consistent theory of quantum gravity. One way to do this is to use axions, which perturbatively have a shift symmetry that protects them from such operators and acquire a potential from instantons and possibly fluxes. In this manner, it is possible in principle to construct a model of natural inflation \cite{Freese:1990rb} or power-law ``axion monodromy'' inflation \cite{McAllister:2008hb}.

In practice, a model of natural inflation requires an axion with super-Planckian decay constant $f > \Mp$, which appears to be difficult to construct in string theory \cite{Banks:2003sx, Rudelius:2014wla, Conlon:2016aea, long:2016jvd, Grimm:2019wtx}. Super-Planckian decay constants are also in tension with the axionic version of the Weak Gravity Conjecture, \cite{ArkaniHamed:2006dz, delaFuente:2014aca, Rudelius:2015xta, Montero:2015ofa, Brown:2015iha}, though possible loopholes to this statement exist.

Super-Planckian decay constants would also violate the bound (\ref{eq:nohilltop}) and produce eternal inflation (under the assumption that the minimum of the axion potential occurs at $V=0$). To see this, we simply expand the natural inflation potential around a local maximum as in (\ref{eq:naturalexpansion}) and plug this into (\ref{eq:nohilltop}) to find the condition for no eternal inflation,
\begin{equation}
f <  \frac{1}{\sqrt{6}} \Mp \approx 0.41 \Mp.
\label{eq:fbound}
\end{equation}
We note that the largest decay constant observed in a $\mathcal{N}=1$ string theory construction (of which we are aware) has $f \approx 0.19 \Mp$, narrowly satisfying (\ref{eq:fbound}).\footnote{\cite{Conlon:2016aea} found an example of a decay constant in the $\mathcal{N}=2$ supersymmetric context with $f \approx 0.52 \Mp$, but the extended supersymmetry in this model precludes a potential of the form (\ref{eq:naturalexpansion}) that would lead to eternal inflation. In the $\mathcal{N}=1$ context, one must worry about moduli stabilization and $\alpha'$ corrections from the small cycle volumes involved in that example, which will likely shrink the allowed values of the decay constant further.} It is also worth noting that models involving multiple axions could satisfy the No Eternal Inflation bound while producing a larger effective decay constant. For instance, consider $N$ identical axions with decay constant $f$ and a potential of the form,
\begin{equation}
V(\bm{\phi}) =  \sum_{i=1}^N  \frac{V_0}{2} \left(1 + \cos\left(\frac{\phi^i}{f}\right) \right).
\end{equation}
In this case, the diagonal direction of the potential, $\phi^i := \phi$, has an effective decay constant of $f_{\rm eff} = \sqrt{N} f$ due to the Pythagorean gain from traveling along the space-diagonal of the $N$-dimensional hypercube \cite{liddle:1998jc, dimopoulos:2005ac}. However, at the maximum of the potential at $\bm{\phi}=0$, we have $V = N V_0$ and $\sum_i \nabla_i \nabla_i V = -N V_0/2f^2$, so at the maximum of the potential,
\begin{equation}
\frac{\sum_{i} \nabla_i \nabla_i V(\bm{\phi})}{V(\bm{\phi})}  = -\frac{1}{2 f^2},
\end{equation}
so the No Eternal Inflation bound (\ref{eq:nohilltop}) implies $f < 1/\sqrt{6} \Mp$, which allows for $f_{\rm eff} \lesssim \sqrt{N/6} \Mp$. In contrast, the Weak Gravity Conjecture for multiple axions \cite{cheung:2014vva} implies $f_{\rm eff}  \lesssim \Mp$ in this model \cite{Rudelius:2014wla}, forbidding a parametric scaling with $N$. Thus, the Weak Gravity bound is strictly stronger than the No Eternal Inflation bound in this context.

In contrast, power-law axion monodromy models are not nearly so tightly constrained by the Weak Gravity Conjecture as natural inflation models \cite{Ibanez:2015fcv,Hebecker:2015zss}, and while one might have to worry about backreaction effects in certain cases \cite{Heidenreich:2015wga, Valenzuela:2016yny, Klaewer:2016kiy, Baume:2016psm, Buratti:2018xjt}, it may well be possible to control these over the $O(10) \Mp$ field range required for a large-field model of inflation. On the other hand, controlling them over a field distance of $O(1000) \Mp$, as required for eternal inflation (\ref{eq:eternalchaotic}), seems much more difficult. It is therefore not unreasonable to think that power-law models of inflation that can be embedded in a quantum gravity theory will not be eternal.

For small-field models, we discussed two possibilities in \S\ref{sec:PRIMORDIAL}: hilltop models and inflection point models. Small-field quadratic hilltop models give rise to eternal inflation but predict $n_s < 0.933$, which is ruled out by experiment. More general hilltop models of degree $p$ seem unlikely due to the required fine-tuning of multiple coefficients in their Taylor expansion, but they can be made to agree with experimental bounds on $n_s$ for $p \geq 6$, and they are also necessarily eternal. Small-field inflection point models yield $n_s > 0.933$ and agree nicely with observation for a range of parameter values, but they do not give rise to eternal inflation. Thus, amongst the seemingly less fine-tuned candidates, observations favor the non-eternal inflection point models over the necessarily eternal quadratic hilltop models.

\subsection{A No Eternal Inflation Principle?}

We have studied the consequences of a No Eternal Inflation principle for models dark energy and inflation, and related them to several other Swampland conjectures. But the most important question remains: why should we believe in a No Eternal Inflation principle in the first place?

The plausibility of such a principle was first advanced in \cite{Page:2006nt} in an attempt to avoid the undesirable prospect of Boltzmann brain domination in an eternally inflation multiverse. Reference \cite{ArkaniHamed:2008ym} subsequently considered this question in more detail, and while the authors ultimately concluded that eternal inflation is more plausibly in the Landscape than the Swampland, they did offer several clues as to why it might be forbidden. First off, exactly stable de Sitter seems problematic for several reasons: it does not allow for precisely measurable observables \cite{Witten:2001kn}, it does not arise in string theory, and it is not possible to increase the lifetime of a metastable de Sitter bubble indefinitely through a Coleman-de Luccia transition. This suggests that de Sitter phases in quantum gravity are likely to be metastable at best.

Assuming this is true, one should presumably think of de Sitter space as a bubble embedded in some larger, asymptotically Minkowski/AdS spacetime, which is defined in terms of S-matrix elements/boundary CFT correlators. But several issues arise if one attempts to do this. As pointed out in \cite{ArkaniHamed:2008ym}, if the de Sitter space is eternally-inflating inside a Minkowski vacuum, its causal structure at late times looks nothing like Minkowski space, so it is not clear how to make sense of the above prescription. Relatedly, one attempt to embed de Sitter in asymptotically Minkowski space found that that it suffered from instabilities \cite{Freivogel:2004rd}. In AdS, the creation of an inflating bubble would correspond in the dual CFT to illegal evolution from a pure state to a mixed state \cite{Freivogel:2005qh}.

Since the publication of \cite{ArkaniHamed:2008ym}, another potential issue with eternal inflation has been discussed in \cite{Fischetti:2014uxa}. Specifically, the authors considered a spacetime consisting of two disconnected, asymptotically AdS regions connected by a dS wormhole. Such spacetimes do not admit codimension-2 extremal surfaces that run from one end of the wormhole to the other, rendering the Hubeny-Rangamani-Takayanagi (HRT) entropy ill-defined. This difficulty disappears if the dS region is regulated so that inflation ends after a finite time.\footnote{Another entropy-related issue arises in the ``bag of gold'' spacetimes constructed by Wheeler \cite{DeWitt:1964oba}: an FRW universe connected by a black hole throat to an asymptotically AdS region would seem to produce an entropy that is far too large from the perspective of an observer outside the black hole \cite{Marolf:2008tx}. However, this paradox applies to general FRW spacetimes, not just eternally-inflating ones, so it does not seem to represent a problem with eternal inflation, specifically.}

A couple of more philosophical points are worth mentioning. First off, as noted in \cite{ArkaniHamed:2008ym}, eternal inflation is sometimes regarded as a necessary ingredient in an anthropic solution to the cosmological constant problem, as it offers a way to populate the string Landscape. However, this is not correct: there may well be other mechanisms for populating the life-permitting vacua in the Landscape, provided that such vacua do exist.\footnote{Some works have even questioned whether or not eternal inflation is capable of populating the Landscape via a fractal-like multiverse, see e.g. \cite{ArkaniHamed:2007ky, Hawking:2017wrd}.} Indeed, it is not even clear that this is a question of physics as opposed to metaphysics.

Secondly, there is no question that a No Eternal Inflation principle would require extraordinary coincidences in particle physics and effective field theory. It is hard to imagine that the decay rate of our own vacuum could be tuned small enough to produce life, yet large enough to prevent eternal inflation. It is harder still to imagine that the decay rate of \emph{every} vacuum in the Landscape could be bounded above in this way, and that \emph{every} point in the effective potential with positive vacuum energy could satisfy the conditions necessary to avoid eternal inflation. These coincidences are probably harder to swallow than any of the problems currently facing eternal inflation, and it is likely that we will have to deal with the physics of eternal inflation whether we like it or not. Nonetheless, it is worth noting that deep, fundamental principles of physics often appear as remarkable coincidences from other (possibly less fundamental) perspectives. For instance, consistency with black hole physics implies statements about the volumes and degeneracies of cycles in Calabi-Yau manifolds that appear remarkable from the perspective of geometry \cite{Vafa:1997gr}. The electron mass seems incredibly fine-tuned in classical electromagnetism without the inclusion of a particle of equal mass and opposite charge (the positron) \cite{Weisskopf:1939zz}, which itself might seem like a remarkable coincidence without the CPT theorem \cite{Schwinger:1951xk}. And Kepler's second law seems like a remarkable coincidence until you learn about Newtonian gravitation. In the present case, there are hints from the holographic perspective that eternal inflation may be in the Swampland. If this is so, the consequences would manifest as great coincidences from the (less fundamental) perspective of effective field theory. More work is needed if this scenario is to be put on firmer footing.

\section{Conclusions}\label{sec:CONC}

We have solved the Fokker-Planck equation describing stochastic inflation, either analytically or numerically, for several simple potentials and used these solutions to derive conditions for eternal inflation. We examined the prospects for eternal inflation in several popular inflationary models. We speculated about the possibility that eternal inflation is in the Swampland, noting similarities between the conditions required to avoid eternal inflation and several proposed Swampland criteria. Finally, we considered phenomenological consequences and possible motivations of a No Eternal Inflation principle.

Our analysis has left us with more questions than answers. Can string theory constructions of de Sitter vacua, quintessence, super-Planckian axion decay constants, and/or critical points with suitably-light tachyonic directions to produce eternal inflation be made more explicit, or can we identify problems with them? Is some version of a de Sitter Conjecture true, and if so, what goes wrong when it is violated? Is eternal inflation incompatible with quantum gravity, and if so, what new physics will ensure that the lifetime of our own universe is short enough to prohibit eternal inflation? We hope these questions can be addressed in the near future, especially if the decay of our vacuum really is imminent.

\section*{Acknowledgements}

We thank David Andriot, Nima Arkani-Hamed, Sven Krippendorf, Liam McAllister, Matthew Reece, Cumrun Vafa, Aron Wall, Edward Witten, Wayne Zhao, and Kathryn Zurek for useful discussions. T.R. is supported by the Carl P. Feinberg Founders Circle Membership and by NSF grant PHY-1606531.

\bibliographystyle{utphys}
\bibliography{ref}

\end{document}